\documentclass[12pt]{article}

\newcommand{\bbr}{I\!\! R}
\newcommand{\bbz}{Z\!\!\! Z}

\newcommand{\x}{arXiv:}
\newcommand{\m}{\mathrm}

\usepackage{graphicx}

\begin{document}
\thispagestyle{empty}
\begin{center}

\null \vskip-1truecm \vskip2truecm {\Large{\bf

\textsf{ Unitarity at Infinity and Topological Holography}

}}

\vskip1truecm {\large \textsf{Brett McInnes}} \vskip1truecm

 \textsf{\\  National
  University of Singapore}

email: matmcinn@nus.edu.sg\\

\end{center}
\vskip1truecm \centerline{\textsf{ABSTRACT}} \baselineskip=15pt
\medskip
Recently it has been suggested that non-gaussian inflationary
perturbations can be usefully analysed in terms of a putative dual
gauge theory defined on the future conformal infinity generated by
an accelerating cosmology. The problem is that unitarity of this
gauge theory implies a strong constraint [the ``Strominger bound"]
on the matter fields in the bulk. We argue that the bound is just a
reflection of the equation of state of cosmological matter. The
details motivate a discussion of the possible relevance of the
``dS/CFT correspondence" to the resolution of the Big Bang
singularity. It is argued that the correspondence may require the
Universe to come into existence along a non-singular spacelike
hypersurface, as in the theories of ``creation from nothing"
discussed by Firouzjahi, Sarangi, and Tye, and also by Ooguri et al.
and others. The argument makes use of the unusual properties of
gauge theories defined on topologically non-trivial spaces.

 \vskip3.5truecm
\begin{center}

\end{center}

\newpage

\addtocounter{section}{1}
\section* {\large{\textsf{1. dS/CFT: Limitations and Applications}}}
Efforts to connect the AdS/CFT correspondence \cite{kn:horpol}
with cosmology lead naturally to the idea of a dS/CFT
correspondence, in which the physics of accelerating spacetimes is
related to a Euclidean CFT defined on spacelike conformal infinity
\cite{kn:strominger}\cite{kn:wittends}. In the form used by
Maldacena \cite{kn:maldacena}\cite{kn:mcnees}\cite{kn:schaar},
this version of holography has recently been revived
\cite{kn:seery} [see also \cite{kn:kachru}], in the hope of
establishing a new understanding of the improved observational
data. There has also been a revival of interest in the theoretical
aspects of the correspondence. Thus for example Polchinski
\cite{kn:polchinski} has recently discussed the idea of
\emph{emergent time} in asymptotically de Sitter spacetimes,
comparing it with ``emergent gauge symmetries" of a possible gauge
theory at future infinity.

There are, however, some doubts as to whether a \emph{precise} de
Sitter analogue of the AdS/CFT correspondence can actually be
constructed. It is generally agreed that a de Sitter phase can only
be metastable in string theory \cite{kn:giddings}\cite{kn:KKLT};
generically it would decay to a cosmological spacetime with a
Crunch; so the future boundary on which the dual theory is to be
defined may not exist. This could simply mean that the dual theory
can only be defined on some spacelike hypersurface at large but
finite proper time, or that the dual theory can only be formulated
in the Euclidean version of the spacetime [as in the work of
Maldacena and Maoz \cite{kn:maoz}], or it may imply more serious
limitations. A concrete proposal for understanding the limitations
of dS/CFT was suggested by van der Schaar \cite{kn:schaar}, who
argued that dS/CFT is effective only at the level of a
coarse-graining in which each lattice site of the field theory
corresponds to an entire static patch in the de Sitter bulk. One way
of stating the case would be to suggest that, on the bulk side of
the correspondence, dS/CFT is primarily relevant to specifically
``cosmological" features of spacetime and its matter content, not to
their detailed structure.

In this spirit, we wish to make an observation regarding a curious
and apparently unphysical feature of the dS/CFT correspondence: it
appears to impose a severe upper bound on the masses of particles.
In detail, the limit of a massive p-form field amplitude [for a
p-form field $\varphi$ of mass m$_{\varphi}$] in de Sitter
spacetime defines a CFT two-point function at de Sitter conformal
infinity. In the p = 0 case, the conformal weight corresponding to
the boundary operator defined by $\varphi$ is given, in four
spacetime dimensions, by
\begin{equation}\label{eq:A}
\m{h_{+}\;=\;{{1}\over{2}}\,[\,3\;+\;\sqrt{9\;-\;4\,L^2\,m_{\varphi}^2}\,]}.
\end{equation}
One sees immediately that the weight will be \emph{complex} unless
m$_{\varphi}$ satisfies
\begin{equation}\label{eq:B}
\m{m_{\varphi}^2 \,\leq\,9/(4L^2)}.
\end{equation}
Presumably a violation of this bound would therefore signal a
failure of unitarity in the field theory. We stress that while
this conclusion is derived for scalar matter, it applies to
\emph{all} kinds of p-form matter: see the discussion of the
analogous extension in the AdS case in \cite{kn:wittenads}.

This ``Strominger bound"
\cite{kn:strominger}\cite{kn:exploring}\cite{kn:cadoni} on
particle masses is analogous to the well-known
Breitenlohner-Freedman bound \cite{kn:freed} on the masses of
p-form fields in anti-de Sitter spacetime; but, being an
\emph{upper} bound, it is at first sight much less reasonable. For
some purposes, it can be ignored: it has been argued by Seery and
Lidsey \cite{kn:seery} that only very low-mass scalars could give
rise to non-trivial perturbations at late times, so the Strominger
bound is not a problem for the recent studies of inflationary
perturbations from the dS/CFT point of view. However, one would
certainly prefer to have a more precise demonstration that the
Strominger bound really does not impose unreasonable conditions on
\emph{any} of the matter fields which are relevant on the scales
to which the dS/CFT correspondence applies.

We shall suggest a solution of this problem inspired by the
arguments of van der Schaar \cite{kn:schaar} mentioned earlier: we
shall assume that dS/CFT should only be expected to yield useful
information at the \emph{cosmological} level, and not on smaller
scales. The relevant forms of matter at this level, apart from the
dark energy/inflaton [both of which we represent by a cosmological
constant] are zero-pressure [non-relativistic] matter, radiation,
and, in the pre-inflationary era, ``fluids" representing strings and
other extended objects. The question is then whether there is some
sense in which these specifically ``cosmological" forms of matter do
in fact satisfy the Strominger bound. If this is \emph{not} the
case, then dS/CFT would imply that the apparently reasonable demand
of unitarity on the part of the field theory imposes unphysical
conditions on bulk physics, and this would call the whole approach
into question.

We argue here that ``cosmological matter" does [barely] satisfy
the Strominger bound: interestingly, zero-pressure matter actually
\emph{saturates} the bound. We conclude that the bound is not
unreasonable, provided that the limits to the validity of dS/CFT,
of the kind pointed out by van der Schaar, are kept in mind. The
agreement of the demand of unitarity on the boundary with the
requirement that the bulk field reproduce the classical equation
of state is rather remarkable.

One of the major hopes for the AdS/CFT correspondence is that it
will allow us to probe the singularities in black holes
\cite{kn:juan}\cite{kn:shenker}\cite{kn:maeda}. Similarly, one
might hope to use AdS/CFT or dS/CFT to probe the Big Bang
singularity \cite{kn:alberghi}. Unfortunately, even in the AdS
case, it is extremely difficult to do this in an even partially
realistic way. For some recent examples: in \cite{kn:veron} an
ingenious attempt is made to study de Sitter space by embedding it
inside an AdS black hole, but the resulting cosmology has no Big
Bang; conversely, the AdS/CFT correspondence has been used to
study Bang singularities in certain cosmological models
\cite{kn:bak}, but these are not asymptotically de Sitter to the
future. For dS/CFT the case is still worse, since so little is
known about the putative boundary theory. In the concluding
section we draw attention to the fact that the topologically
non-trivial spacetime structure assumed here [we take the spatial
sections to be tori] can be expected to give rise to some
unfamiliar physics, of the kind discussed some time ago by Krauss
and Wilczek \cite{kn:wilczek} and by Preskill and Krauss
\cite{kn:preskill}. If the field theory at infinity is of this
kind, in which parallel transport around non-contractible loops
can cause discrete transformations, then this can give rise to
very puzzling behaviour deep in the bulk, near to the initial
singularity. We speculate that this kind of ``topological
holography" could have some role to play in resolving that
singularity, when the dS/CFT correspondence is better understood.

\addtocounter{section}{1}
\section*{\large{\textsf{2. Representing Cosmological Matter by a Massive Field }}}
In this section, we show how to represent the familiar forms of
cosmological matter in a way that can be connected with the
Strominger bound. At the same time, we must incorporate the effect
of such matter on the spacetime geometry.

We begin with the version of de Sitter spacetime with flat spatial
sections; we compactify these sections to copies of the [cubic]
three-torus T$^3$. It has been argued elsewhere that this is the
most suitable global structure for investigations of quantum
cosmology
\cite{kn:tallandthin}\cite{kn:singularstable}\cite{kn:OVV}. One of
the many virtues  of this toral version of de Sitter spacetime is
that its boundary at infinity is \emph{compact and connected}, so
that we can avoid the subtle issues \cite{kn:strominger} which arise
when the dS/CFT correspondence is constructed on the spatially
spherical version of de Sitter spacetime, which has two spheres at
infinity. Here, instead, we have \emph{one} torus at [future]
infinity. Henceforth, then, our working hypothesis is that the
spacetime topology is $\bbr\,\times\,\m{T}^3$ [or possibly some
non-singular quotient].

Thus the metric, before we introduce any matter, is that of
Spatially Toral de Sitter:
\begin{equation}\label{eq:C}
g(\m{STdS})\;=\;\m{dt^2\; -\;K^2\,e^{(2\,t/L)}\,[\,d\theta_1^2 \;+\;
d\theta_2^2 \;+\; d\theta_3^2]},
\end{equation}
where $\theta_{1,2,3}$ are angular coordinates on a cubic
three-torus, where K defines the spatial length scale, L is the
scale defined by the cosmological constant, and we use
($+\;-\;-\;-$) signature.

Now we wish to introduce into this spacetime the familiar
cosmological matter fields --- non-relativistic matter, radiation,
cosmic strings and so on. A common property of all these forms of
matter is that they have a \emph{constant} equation-of-state
parameter w, where w specifies the ratio of pressure to density in
the homogeneous case. In order to determine whether the Strominger
bound is satisfied, we have to find a formal representation of these
matter fields by some p-form field with a definite mass. In view of
the isotropy and homogeneity of cosmological matter, the obvious
choice is to try to do this using a \emph{scalar} field [p = 0].

The solution is as follows. We take STdS spacetime, and introduce
into it a homogeneous scalar field $\varphi$ with the usual kinetic
term and with a potential
\begin{equation}\label{eq:J}
\m{V(\varphi,\;\epsilon)\;=\;-\,{{3}\over{8\pi
L^2}}\,[1\;-\;{{1}\over{6}}\,\epsilon]\,sinh^2(\sqrt{2\,\pi\,\epsilon}\,\varphi)};
\end{equation}
here $\epsilon$ is a positive constant. Bear in mind that STdS
spacetime in this signature has a \emph{negative} cosmological
constant $-\,3/$L$^2$ which gives rise to a positive energy density
$3/(8\pi$L$^2$). The precise numerical factor in the potential is
chosen for later convenience\footnote{This choice normalizes the
cosmological scale factor in a convenient way.}.

We now make the following claim: we assert that, as far as the
spacetime geometry is concerned, the field $\varphi$ \emph{exactly
mimics} the effects on STdS spacetime of a homogeneous matter field
with positive energy density and a \emph{constant} equation-of-state
parameter related to $\epsilon$ by
\begin{equation}\label{eq:K}
\m{w_{\varphi}\;=\;{{1}\over{3}}\,\epsilon\;-\;1}.
\end{equation}
Thus for example if we insert non-relativistic matter [zero
pressure, hence w = 0] into the STdS spacetime and allow it to act
on the spacetime geometry, this will have the same effect as
introducing $\varphi$ with $\epsilon$ = 3, while $\varphi$ with the
value $\epsilon$ = 4 mimics the effects of radiation; $\epsilon$ = 1
corresponds to a static network of planar domain walls, increasing
to 1.5 for the scaling regime \cite{kn:domain}; $\epsilon$ = 2 for a
static network of strings, and so on. We stress that we are
\emph{not} primarily interested in using this field to violate the
Strong Energy Condition --- asymptotically at least, the
acceleration is due to the negative contribution made by the STdS
cosmological constant to the total pressure. Instead, $\varphi$ just
represents any form of homogeneous cosmological matter, with a
constant equation-of-state parameter $\m{w}_{\varphi}\;\geq\;-\,1$,
which is to be superimposed on the STdS spacetime.

We now proceed to justify these claims. We shall consider
Friedmann cosmological models with metrics of the form
\begin{equation}\label{eq:ALPHA}
g \;=\; \m{\m{dt}^2\; -\; \m{K^2\;a(t)}^2[d\theta_1^2 \;+\;
d\theta_2^2 \;+\; d\theta_3^2]};
\end{equation}
this generalizes the STdS metric in an obvious way. Adding the
energy density of the $\varphi$ field to that of the initial STdS
space, we have a Friedmann equation of the form
\begin{equation}\label{eq:BETA}
\m{\Big({{\dot{a}\over{a}}}\Big)^2 \;=\;
{{8\pi}\over{3}}\;\Big[\;{{1}\over{2}}\;\dot{\varphi}^2
\;-\;\textup{V}(\varphi,\;\epsilon) \;+\;{{3}\over{8\pi L^2}}\Big]}.
\end{equation}
The equation for $\varphi$ itself is
\begin{equation}\label{eq:GAMMA}
\m{\ddot{\varphi}\;+\;3\;{{\dot{a}}\over{a}}\;\dot{\varphi}\;-
\;{{dV(\varphi,\;\epsilon)}\over{d\varphi}}\;=\;0.}
\end{equation}
Surprisingly, these equations have very simple solutions: one
finds that [with natural initial conditions] $\varphi$ is given by
\begin{equation}\label{eq:DELTA}
\m{\varphi\;=\;{{1}\over{\sqrt{\pi\epsilon/2}}}\,tanh^{-\,1}(e^{-\,\epsilon\,t/2L})},
\end{equation}
and the metric is\footnote{The reader who wishes to undertake the
task of verifying these solutions will find the following simple
fact helpful: if A and B are quantities related by tanh(A) =
e$^{-\m{B}}$, then cosh(2A) = coth(B).}
\begin{equation}\label{eq:EPSILON}
g\m{(\epsilon,\,K,\,L) \;=\; dt^2\; -\;
K^2\;sinh^{(4/\epsilon)}\Big({{\epsilon\,t}\over{2L}}\Big)\,[d\theta_1^2
\;+\; d\theta_2^2 \;+\; d\theta_3^2]}.
\end{equation}
From these results one can compute the energy density and pressure
of the $\varphi$ field alone:
\begin{equation}\label{eq:ZETA}
\m{\rho_{\varphi}\;=\;{{3}\over{8\pi
L^2}}\,cosech^2\Big({{\epsilon\,t}\over{2L}}\Big)},
\end{equation}
\begin{equation}\label{eq:ETA}
\m{p_{\varphi}\;=\;{{3}\over{8\pi
L^2}}\,\Big[\,{{1}\over{3}}\,\epsilon\;-\;1\,\Big]\,cosech^2\Big({{\epsilon\,t}\over{2L}}\Big)},
\end{equation}
from which equation (\ref{eq:K}) above is immediate. Note that the
density decays more rapidly with increasing time for larger values
of $\epsilon$.

Since we are using the standard kinetic term, $\varphi$
automatically satisfies the Null Energy Condition. To see the
conditions under which the Strong Energy Condition is satisfied, we
have to determine whether the quantity $\Pi$ = $\rho$ + 3p is
non-negative, where $\rho$ and p are the \emph{total} energy density
and total pressure [including the contributions due to the
background cosmological constant, which total to $-\,3/4\pi$L$^2$]
respectively. For the metrics we are concerned with here, $\Pi$ is a
function of $\epsilon$, L, and t given by
\begin{equation}\label{eq:STRONG}
\m{\Pi(\epsilon,\,L,\,t)\;=\;{{-\,3}\over{4\pi
L^2}}\;+\;\rho_{\varphi}\;+\;3\,p_{\varphi}\;=\;{{-\,3}\over{4\pi
L^2}}\,\Big[\,1\;+\;(1\;-\;{{\epsilon}\over{2}})\,cosech^2\Big({{\epsilon\,t}\over{2L}}\Big)\Big].}
\end{equation}
We see that the SEC is always violated eventually, but that it is
satisfied in the early Universe \emph{provided} that
$\epsilon\;>\;2$; \emph{otherwise it is violated at all times}, just
as it is in de Sitter spacetime. In physical terms, this means that,
for all $\epsilon\;>\;2$ the Universe \emph{decelerates} in its
earlier [post-inflationary] stages, and only later begins to
accelerate: this is the case for the post-inflationary phase of our
Universe. For $\epsilon\;\leq\;2$, by contrast, the spacetime always
accelerates, and in fact for $\epsilon\;<\;2$ the acceleration
actually \emph{diverges} as t = 0 is approached. We shall have more
to say about this below.

If $\epsilon$ = 3, we should have the local metric for a spacetime
containing non-relativistic matter and a de Sitter cosmological
constant, and indeed $g\m{(3,\,K,\,L)}$ gives
--- purely locally --- the classical Heckmann metric
[see \cite{kn:overduin} for a recent discussion]. In the general
case it agrees [again locally] with the results reported in
\cite{kn:eroshenko}, where it is obtained by postulating a linear
equation of state [without giving a matter model]. For large t we
have
\begin{equation}\label{eq:THETA}
g\m{(\epsilon,\,2^{\,2/\epsilon}\,K,\,L) \;\approx\; dt^2\;
-\;K^2\,e^{2\,t/L}[d\theta_1^2 \;+\; d\theta_2^2 \;+\;
d\theta_3^2]},
\end{equation}
which is the STdS metric given in equation (\ref{eq:C}); notice that
$\epsilon$ effectively drops out. Thus our metric is
``asymptotically STdS", for all $\epsilon$.

All of the metrics in (\ref{eq:EPSILON}) \emph{appear} to be
singular at t = 0, but we remind the reader that such appearances
can be deceptive: for example, consider the metric
\begin{eqnarray}\label{eq:OLIPHAUNT}
g(\mathrm{SHdS_4})\; =\;
\m{dt}^2\;-\;\m{sinh^2(t/L)}\,[\mathrm{dr^2\; +\;\m{L^2}\,
sinh^2(r/L)}\{\mathrm{d}\theta^2 \;+\;
\mathrm{sin}^2(\theta)\,\mathrm{d}\phi^2\}],
\end{eqnarray}
with hyperbolic spatial sections [which we do \emph{not} compactify
here]. This appears to be singular in the same way, but in fact this
is just a version of de Sitter spacetime, the [non-compactified]
``spatially hyperbolic" version \cite{kn:OVV}, which is, of course,
entirely non-singular. The question as to which members of
(\ref{eq:EPSILON}) are really singular is important, and we shall
return to it below.

We have shown how to represent cosmological matter in a way that is
relevant to the Strominger bound. Let us now show that the bound is
in fact satisfied by such matter.

\addtocounter{section}{1}
\section*{\large{\textsf{3. Cosmological Matter Satisfies the Strominger Bound }}}

At late times, the metric $g\m{(\epsilon,\,K,\,L)}$ [equation
(\ref{eq:EPSILON})] is locally indistinguishable from that of de
Sitter spacetime, and, furthermore, $\varphi$ is very small
[equation (\ref{eq:DELTA})]; hence we see, from (\ref{eq:J}), that
$\varphi$ can be regarded as a scalar field of squared mass
\begin{equation}\label{eq:NU}
\m{m_{\varphi}^2\;=\;{{3}\over{2L^2}}\,\epsilon\,[1\;-\;{{1}\over{6}}\,\epsilon]}
\end{equation}
propagating on a local de Sitter background. By computing
\begin{equation}\label{eq:ANGOLA}
\m{
{{9}\over{4L^2}}\;-\;m_{\varphi}^2\;=\;(\epsilon\;-\;3)^2/4L^2\;\geq\;0},
\end{equation}
we see at once that the Strominger bound is automatically ---\emph{
but only barely} --- satisfied for all values of $\epsilon$. The
conformal weight is real: from (\ref{eq:A}) we have
\begin{eqnarray} \label{eq:OMICRON}
\m{h}_{+} & = & \epsilon/2,\;\;\epsilon\;\geq\;3   \nonumber \\
          & = & 3\;-\;(\epsilon/2),\;\; \epsilon\;\leq\;3.
\end{eqnarray}
Notice that by varying $\epsilon$, either from 0 to 3 or from 3 to
6, one can obtain all values of the weight allowed by the Strominger
bound [that is, all values between 3/2 and 3]: in fact, apart from
3/2, all allowed values of the weight can be obtained \emph{in two
different ways}. Thus for example h$_+$ = 2 can be obtained either
by choosing $\epsilon$ = 2 or by taking $\epsilon$ = 4.

From (\ref{eq:ANGOLA}) we see that the Strominger bound is most
nearly violated when $\epsilon$ = 3; it is saturated in this case,
that is, \emph{in the case of zero-pressure, non-relativistic
matter.} This has a deep mathematical significance, as follows.
Because de Sitter spacetime and Euclidean hyperbolic space have the
same isometry group, the relevant representation theory has been
extensively developed \cite{kn:vilenkin}\cite{kn:gazeau}. The scalar
representations fall into three families, the principal,
complementary, and discrete series. The principal representations
are those which, under contraction of the de Sitter group to the
Poincar\'e group, correspond to the familiar flat space
representations. They are of two kinds, which in fact are classified
precisely by the weight h$_+$ which appears in the dS/CFT
correspondence. The first kind is the case where h$_+$ is complex:
these of course violate the Strominger bound. The only other
principal representation corresponds precisely to the case where the
bound is saturated.

The complementary series consists of representations where the
Strominger bound is satisfied but not saturated\footnote{The
relevant Hermitian form in the complementary case
[\cite{kn:vilenkin}, page 518] involves a gamma function which is
ill-defined if the Strominger bound is saturated, so the $\epsilon$
= 3 representation certainly does not belong to this series.}; these
representations have no flat-spacetime analogue. [The discrete
series corresponds to the special, massless case.] We conclude that
$\epsilon$ = 3 is the \emph{only} value which corresponds to a
well-defined flat space representation. It is interesting that
$\epsilon$ = 3 is singled out in this way, for this corresponds to
the kind of matter that, in classical cosmology, is the most
important at late times. It is also interesting that the field we
have used to represent asymptotically de Sitter cosmology with
parameter $\epsilon$ corresponds to precisely the \emph{same} de
Sitter representation as the cosmology with parameter 6 $-$
$\epsilon$.

We conclude that the Strominger bound is in fact satisfied by all
kinds of ``cosmological matter". We have shown this by
representing cosmological matter by fields which, except in the
case of non-relativistic matter, correspond exactly to the
complementary series of representations of the de Sitter group.
The perfect agreement here --- note the fact that the most
familiar kind of cosmological matter just exactly saturates the
bound --- is rather remarkable, in that the Strominger bound is
required by a \emph{quantum} condition on the boundary [unitarity
of the field theory] while the form of the potential (\ref{eq:J})
is dictated by the demand that the \emph{classical} equation of
state of the bulk matter should be reproduced\footnote{In this
regard it may be of interest to examine the status of the
Strominger bound when more complex equations of state are
considered; see for example \cite{kn:od}.}.

Our discussion shows that the cosmological models with metrics
given by equation (\ref{eq:EPSILON}) do define a well-behaved CFT
at infinity. Since these models are both asymptotically de Sitter
and [unlike de Sitter spacetime itself] have large energy
densities and pressures in the remote past, they are more
realistic than de Sitter spacetime, and they are the natural
models to consider if one hopes to use the dS/CFT correspondence
to try to understand the distant past of the Universe. We conclude
with some observations regarding these spacetimes and the possible
relevance of dS/CFT to the problem of the initial singularity.

\addtocounter{section}{1}
\section*{\large{\textsf{4. Comments on The Initial Singularity and dS/CFT}}}
The recent work of Seery and Lidsey \cite{kn:seery} is based on the
idea that the dS/CFT correspondence may be able to teach us
something about the inflationary era. It is natural to extend this
idea to ask whether the correspondence may be brought to bear on the
\emph{pre}-inflationary era, about which very little is known. The
most important problem which arises here is that of the initial
singularity, a problem which is not solved by Inflation
\cite{kn:borde}. In the case of AdS/CFT, good behaviour [such as
unitarity] on the part of the boundary field theory can be used to
argue \cite{kn:juan} for good behaviour in the bulk [avoidance of
information loss in black hole evaporation]. Ultimately one hopes to
construct an analogous argument in the dS/CFT case.

Unfortunately, very little is understood about the field theory at
infinity in this case, so we cannot yet hope to replicate the
progress that has been made in the AdS case. However, the spacetimes
we have been discussing do have some unusual properties which may
suggest novel approaches to the problem of the initial singularity.

First let us consider the field theory on the boundary. The key
point to note here is that the conformal boundary is geometrically
very simple [it is flat] but \emph{topologically non-trivial}: it is
either a torus or perhaps some non-singular quotient of a torus
\cite{kn:conway}\cite{kn:reallyflat}. The fundamental group is
infinite, that is, there are non-contractible curves which wind
around the space arbitrarily many times\footnote{The boundary can of
course have a non-trivial fundamental group even in the locally
spherical case, but this group is always finite in that case.}.

Gauge theories on such spaces can have a number of very interesting
properties, going far beyond the need to impose periodic boundary
conditions. For example, extending the work of Krauss and Wilczek
\cite{kn:wilczek} on the theory of ``discrete gauge hair" on black
holes\footnote{See \cite{kn:coleman} for a comprehensive review of
this subject.}, Preskill and Krauss \cite{kn:preskill} were led to
study the extraordinary phenomena arising when gauge theories with
\emph{discrete or disconnected} gauge groups are constructed on
spaces with non-trivial fundamental groups. These occur very
naturally when gauge groups are embedded in larger groups, as in
Grand Unified theories. Let us take a concrete example, though we
stress that phenomena of the kind we are about to describe can occur
in very many other ways.

When the SU(3) of QCD is embedded in the Spin(10) Grand Unified
Group [usually known, not quite accurately, as ``SO(10)"], one finds
that there are many disconnected subgroups of Spin(10) with SU(3) as
identity component. An interesting example of such a subgroup has
the following form: it may be expressed as a \emph{semi-direct}
product of $\bbz_4$ with SU(3):
\begin{equation}\label{eq:ICELAND}
\m{SU(3)}\;\triangleleft\;\bbz_4\;=\;\m{SU(3)\;\bigcup\;\gamma
\,\cdot\,SU(3)\;\bigcup\;(-\,1)\,\cdot\,SU(3)\;\bigcup\;(-\,\gamma)\,\cdot\,SU(3),}
\end{equation}
where $-\,1$ is the element of Spin(10) which is factored out to
obtain SO(10), and where $\gamma$ is a certain Spin(10) element with
$\gamma^2$ = $-\,1$, such that conjugation by $\gamma$ has the
effect of complex conjugation on SU(3). Conjugation by $\gamma$ also
has the effect of complex conjugation on the electromagnetic U(1)
embedded in Spin(10); hence it is related to charge conjugation.

Such disconnected groups are of physical interest if one can prove
that there is a gauge connection with a \emph{holonomy group} ---
the group generated by gauge parallel transport around closed loops
--- which contains an element representing $\gamma$. It is in fact possible to prove
\cite{kn:alice} that, with topology T$^3$, there do exist
$\m{SU(3)}\;\triangleleft\;\bbz_4$ gauge connections with holonomy
groups isomorphic to the $\bbz_4$ generated by $\gamma$. This is
analogous to parallel transport on a [flat] Klein bottle: despite
the fact that the metric is flat, parallel transport around
certain non-contractible loops on the Klein bottle can give rise
to a transformation which reverses orientation. The [linear]
holonomy group of the Klein bottle is just $\bbz_2$, generated by
an element of the orthogonal group O(2) which does not lie in the
connected component of the identity.

If we had a gauge connection on the torus at infinity with
holonomy group $\bbz_4$ generated by $\gamma$ as above, then
moving an object around certain closed, non-contractible curves
would cause particles to transform to anti-particles, and vice
versa: the system is ``non-orientable" with respect to charge
conjugation instead of parity. The very remarkable properties of
such gauge theories are described in detail by Preskill and Krauss
in \cite{kn:preskill}, to which the reader is referred for further
details. Here we shall not need to consider a specific gauge group
like $\m{SU(3)}\;\triangleleft\;\bbz_4$: all we need is to
understand that we should expect parallel transport around certain
non-contractible loops on the torus to implement \emph{discrete}
transformations, analogous to [but probably different from] charge
conjugation.

There are three specific points to be made here. First, with spatial
topology T$^3$, there is nothing artificial about gauge connections
with holonomy groups of this kind; indeed, such behaviour should be
regarded as generic when gauge theories are constructed on the
torus, provided that a disconnected gauge group arises. Second, in
the cosmological context, one might require a gauge field to have
vanishing field strengths; but phenomena of the kind discussed by
Preskill and Krauss \emph{can still be present} even in this case,
in the manner of the Aharonov-Bohm effect\footnote{See
\cite{kn:rachel} for a recent discussion of such effects in
cosmology.}. Third, such gauge fields are in no sense pathological;
that is, we have no reason whatever to expect any kind of unphysical
behaviour on the part of the boundary field theory, though the
details may be unfamiliar
--- for example, conservation of charge [or whatever conservation
law is relevant to the discrete symmetry in question] is enforced in
a very unusual way, through ``Cheshire charges" \cite{kn:preskill}.
In fact, non-trivial holonomies of this kind appear routinely in
string theory in the form of ``Wilson lines", which have been used
recently in string cosmology \cite{kn:quevedo}.

We conclude that, if anything resembling the AdS/CFT correspondence
is valid here, then it should be possible to construct a model of
bulk physics which remains non-pathological even when the boundary
theory fully incorporates effects due to the toral topology. As we
shall see, this leads us in a very interesting direction.

In order to give a more concrete basis for discussion, let us
construct a very simple explicit model of the pre-inflationary
era. In fact, some of the spacetimes discussed above are ideally
suited to this. For all of them evolve naturally to an
``inflationary" state, that is, they are all asymptotically de
Sitter to the future, and they all have a [single] torus as their
conformal boundaries. On the other hand, some of them can be used
to set up a semi-realistic model of conditions in the very early
Universe, which may well have contained networks of extended
objects of the kind encountered in, for example, string gas
cosmology \cite{kn:bat}. Such networks correspond to matter fields
with energy densities that decay relatively slowly; in fact, using
equation (\ref{eq:K}), one can show that the relevant range of
$\epsilon$ is $\epsilon\;\leq\;2$. We stress that it is the
\emph{pre}-inflationary era that is being discussed here; the
usual lore regarding the cosmology of extended objects is not
relevant.

For example, take the case of a spacetime of topology
$\bbr\,\times\,\m{T}^3$ containing dark energy and a static network
of planar domain walls, described by (\ref{eq:EPSILON}) with
$\epsilon\;=\;1$:
\begin{equation}\label{eq:EPSILON1}
g\m{(1,\,K,\,L) \;=\; dt^2\; -\;
K^2\;sinh^4\Big({{t}\over{2L}}\Big)\,[d\theta_1^2 \;+\; d\theta_2^2
\;+\; d\theta_3^2]}.
\end{equation}
This \emph{appears} to be singular at t = 0, but, as we mentioned
earlier, this could be deceptive. In the metric given in
(\ref{eq:OLIPHAUNT}), what is contracting to zero size is not a
spatial section, but rather the separation of neighbouring members
of the congruence of timelike curves which defines the given
coordinates. The same might happen here. Physically, one can argue
very strongly that the metric in (\ref{eq:EPSILON1}) should
\emph{not} be singular, as follows.

The metrics in (\ref{eq:EPSILON}) with $\epsilon\;>\;2$ \emph{are}
in fact singular; but this is just what we would expect. For we
saw [equation (\ref{eq:STRONG})] that in those cases the Strong
Energy Condition holds in the early Universe, and a singularity is
then demanded by the classical singularity theorems. In physical
terms, gravity is attractive when the SEC holds, and so the
presence of a singularity is to be expected. [Since the boundary
theory is assumed to be well-behaved, this presumably just means
that such matter fields are not relevant in the very early
Universe --- which is reasonable.] By contrast, the dark energy
corresponding to a de Sitter cosmological constant generates
gravitational repulsion, and this is how we understand the fact
that pure de Sitter spacetime is \emph{not} singular.

By this logic, it would not make sense for the spacetime with metric
$g\m{(1,\,K,\,L)}$ to be singular: here, gravitation is
\emph{always} repulsive, just as it is in the case of the de Sitter
spacetime [equation (\ref{eq:C})] of which it is a deformation. For
the latter, the quantity $\rho$ + 3p, which [when negative] measures
the extent of violation of the Strong Energy Condition, is given by
the negative constant $-\;3/4\pi$L$^2$; whereas here [from
(\ref{eq:STRONG})] we have
\begin{equation}\label{eq:STRONG1}
\m{\Pi(1,\,L,\,t)\;=\;{{-\,3}\over{4\pi L^2}}\,\Big[\,
1\;+\;{{1}\over{2}}\,cosech^2 \Big({{t}\over{2L}}\Big)\Big].}
\end{equation}
Evidently the function $\m{\Pi(1,\,L,\,t)}$ is always negative, and
in fact it is more negative than in the case of de Sitter spacetime.
As this quantity measures ``gravitational repulsion", there is in
fact even less reason for the spacetime to be singular here than in
the de Sitter case. It is true that $\m{\Pi(1,\,L,\,t)}$ tends to
negative infinity as t approaches zero, but this could be due to a
singularity of the congruence of timelike curves being used here,
and not of the spacetime. We saw that this accounted for the
apparent singularity in the metric in equation (\ref{eq:OLIPHAUNT}).
Our discussion leads us to expect something similar here.

It is therefore very surprising to find that $g\m{(1,\,K,\,L)}$
\emph{is in fact singular}. The scalar curvature of this metric is
\begin{equation}\label{eq:IZOLA}
\m{R}(g\m{(1,\,K,\,L)) \;=\;
-\,{{12}\over{L^2}}\;-\;{{9}\over{L^2}}\,cosech^2\Big({{
t}\over{2L}}\Big)},
\end{equation}
so indeed we have a curvature singularity at t = 0; in fact, all of
the metrics with $\epsilon\;\leq\;2$ are singular. To appreciate
fully what this means, take this metric and define it on the
interval ($-\,\infty,\;0$), so that it describes a contracting
cosmology. The ``force of gravitational repulsion" \emph{increases
without limit} as the Universe contracts, and yet it manages to
contract to zero size. This appears to be self-contradictory
behaviour.

One's first reaction to this extraordinary situation is to suggest
that the singularity is a result of the special boundary conditions
or special symmetries of the metric, and that a more realistic model
would ``bounce", just as the spatially spherical version of de
Sitter spacetime does, and for the same reason. \emph{This is not
correct, however}: Andersson and Galloway
\cite{kn:andergall}\cite{kn:gall}\footnote{See
\cite{kn:singularstable} for a discussion.} have proved results
which essentially imply that that \emph{every} future asymptotically
de Sitter metric defined on a manifold of topology
$\bbr\,\times\,\m{T}^3$ must be geodesically incomplete in the past
if the Null Energy Condition and the Einstein equations hold at all
times. Thus boundary conditions and symmetries cannot be ``blamed"
for the fact that $g\m{(1,\,K,\,L)}$ is singular.

The next suggestion is that the metrics with $\epsilon\;\leq\;2$
might perhaps be ruled out by some internal contradiction in dS/CFT
itself, and indeed it was this possibility that inspired the present
investigation. But we have seen that the relevant forms of matter
define a well-behaved unitary field theory at infinity; matter with,
for example, $\epsilon$ = 2 is precisely as far from violating
unitarity at infinity as ordinary radiation [$\epsilon$ = 4], and
indeed both are represented formally in our analysis by fields in
the same representation of the de Sitter group. Hence there is no
hint that the spacetimes with $\epsilon\;\leq\;2$ are unacceptable
from a holographic point of view. Furthermore, it has been shown
\cite{kn:singularstable} that these spacetimes are stable against
known \cite{kn:porrati} perturbative and non-perturbative
instabilities in string theory.

The Andersson-Galloway results imply that it is the \emph{topology}
of conformal infinity that is to be ``blamed" for the singularities
in these metrics, \emph{as long as the Einstein equations are
assumed to hold everywhere}. But we have stressed that, while the
topology of the boundary can be expected to influence the boundary
gauge theory in important ways, it certainly should not give rise to
any unphysical effects. We conclude that if anything like a dS/CFT
correspondence is valid, the bulk singularity in the metrics with
$\epsilon\;\leq\;2$ will be resolved, and this will happen because
\emph{the boundary topology} will enforce a suitable modification of
the Einstein equations\footnote{The only alternative is a violation
of the Null Energy Condition, which we consider to be implausible;
see \cite{kn:singularstable}\cite{kn:crem}\cite{kn:bun} for
discussions of this.}. This is certainly a reasonable prediction
since, as we saw, it is not physically reasonable for the metrics
with $\epsilon\;\leq\;2$ to be singular in the first place.

The boundary topology can have physical consequences in a variety
of ways, but we wish to propose that the strange effects discussed
by Preskill and Krauss \cite{kn:preskill}, which directly link the
boundary topology to gauge theory physics, are relevant here. How
can these effects enforce a non-singular spacetime structure? Only
a much more detailed understanding of the correspondence can tell
us exactly how this is accomplished, but a further examination of
the metrics in (\ref{eq:EPSILON}) does reveal some striking hints.

Let us assume that that the field theory at infinity has a
discrete or disconnected gauge group, so that parallel transport
around non-contractible loops can impose some discrete
transformation. For the sake of clarity we shall refer to this
discrete transformation as ``charge conjugation", but we stress
that charge conjugation is \emph{just an example} of the kind of
transformation that can arise in this way.

If a dS/CFT correspondence is valid, we can expect a global
``particle/antiparticle" ambiguity at infinity to be inherited by
the spatial sections of the bulk. At the present time, and indeed
by the end of Inflation, this would be physically irrelevant,
since circumnavigations of the Universe along non-contractible
curves are of course by that time not possible. \emph{They may
well, however, have been possible in the pre-inflationary era},
and this we now investigate.

There is in fact a major difference between the metrics of the form
$g\m{(\epsilon,\,K,\,L)}$ with $\epsilon\;\leq\;2$ [which we
originally hoped to use to describe the pre-inflationary era] and
those with $\epsilon\;>\;2$; this may be seen as follows. As
sinh$\m{^{(2/\epsilon)}({{\epsilon\,t}\over{2L}}})$ is approximately
proportional to t$^{(2/\epsilon)}$ when t is small, it follows that
the Lorentzian [angular] conformal time $\eta$ [defined by d$\eta$ =
dt/Ksinh$\m{^{(2/\epsilon)}({{\epsilon\,t}\over{2L}}})$] converges
as t tends to zero, \emph{provided} that $\epsilon\,>\,2$. On the
other hand, if t is large,
sinh$\m{^{(2/\epsilon)}({{\epsilon\,t}\over{2L}}})$ resembles a
multiple of e$\m{^{t/L}}$, and so $\eta$ always converges as t tends
to infinity. Since the spatial sections are always finite, it
follows that, in the case $\epsilon\,>\,2$, our spacetime is
conformal to a piece of Minkowki space which is finite in both space
and time.

\begin{figure}[!h]
\centering
\includegraphics[width=0.6\textwidth]{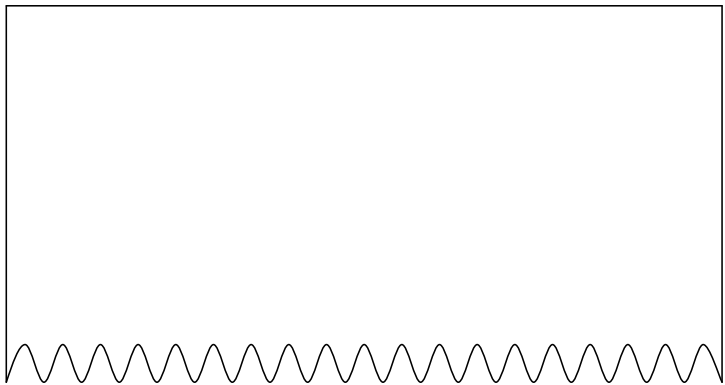}
\caption{Penrose diagram for
$g\m{(\epsilon,\,K,\,L)}\;\m{spacetime}, \;\epsilon\,>\,2$.}
\end{figure}

The Penrose diagram for this case is shown in Figure 1. Future
conformal infinity is spacelike, and there is a Big Bang singularity
which is also spacelike. The horizontal direction represents only
half of the range of any angular coordinate on the torus, namely the
half corresponding to angles between zero and $\pi$. Thus the
vertical lines represent the origin and the point which in this
direction is most distant from the origin. [The reader may find it
helpful to compare this with the diagrams in \cite{kn:chiba}.]

The width of the diagram is given by $\pi$. Its height, in the
$\epsilon\,>\,2$ case, is given by
\begin{equation}\label{eq:ITHACA}
\m{H(L/K,\;\epsilon)\;=\;{{2\,L}\over{K\epsilon}}\,\int_0^{\infty}\,{{dx}\over{sinh^{(2/\epsilon)}(x)}}},
\end{equation}
that is, it is proportional to L/K, and it is related to $\epsilon$
in a more complicated way. For example, for $\epsilon$ = 3
[non-relativistic matter], the height is approximately
$2.81\,\times\,$L/K.

The shape of this diagram determines whether the Universe can be
circumnavigated. For example, with the choice of parameters
leading to the particular shape shown in Figure 1, it is clear
that no particle or signal can circumnavigate the spatial torus,
even in the infinite proper time available. However, if we allow
$\epsilon$ to descend towards 2, the integral in equation
(\ref{eq:ITHACA}) becomes steadily larger. For values just above
2, H(L/K, $\epsilon$) will be much larger than $\pi$, even if L
and K are of similar magnitudes. In that case, the Penrose diagram
will resemble Figure 1, but it will be much taller than it is
wide. In this case, circumnavigation is easy; in fact, for very
tall diagrams, a generic worldline will wrap around the
torus\footnote{Recall again that we are dealing with the situation
\emph{before} Inflation; by the end of Inflation,
circumnavigations will no longer be possible.}.

Circumnavigations of the early Universe can be of interest for a
variety of reasons. For example, they can help to preserve any
initial homogeneity, a fact which is the basis of Linde's model of
low-scale Inflation \cite{kn:lindetypical}\cite{kn:lindenew}. They
are also of interest to us here, however, because we are assuming
that circumnavigations can convert particles to anti-particles [or
cause some other similar discrete transformation]. Suppose that
$\epsilon$ is just above 2, so that the Penrose diagram is much
taller than it is wide; then a generic timelike worldline,
representing the history of [say] a quark, will execute a
definite, non-zero but \emph{finite} number of circumnavigations
as we trace it back to the singularity, and so we can specify
whether this particle emerged from the singularity as a quark or
as an antiquark. But now let us ask what happens when
$\epsilon\;\leq\;2$, as we wish to assume here.

\begin{figure}[!h]
\centering
\includegraphics[width=0.6\textwidth]{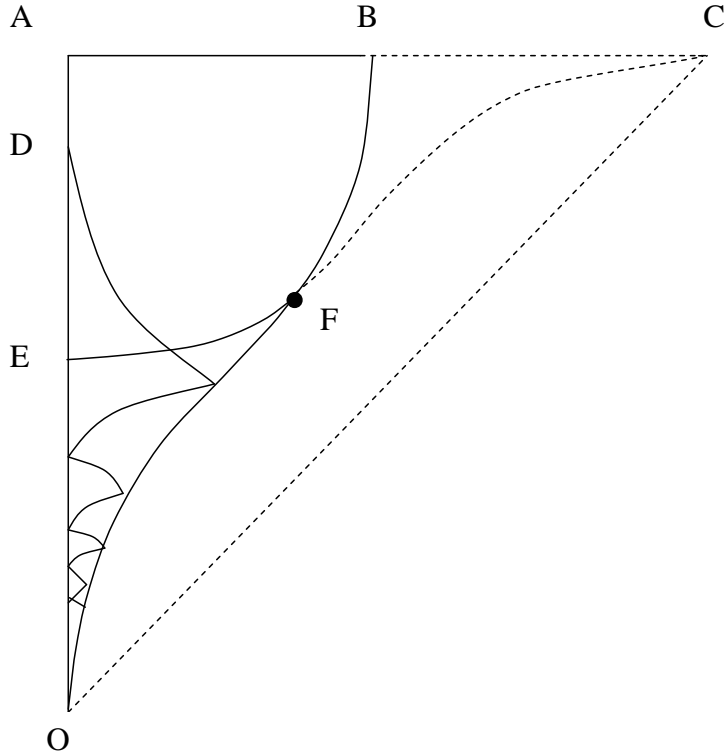}
\caption{Penrose diagram for
$g\m{(\epsilon,\,K,\,L)}\;\m{spacetime}, \;\epsilon\,\leq\,2$.}
\end{figure}

If $\epsilon\;\leq\;2$, then conformal time still converges as t
tends to infinity, but it diverges as t tends to zero. In this case,
the Penrose diagram has an unusual structure pictured in Figure 2.
The triangle OAC represents the lower half of the Minkowski
conformal diagram, and the line EFC represents a typical spacelike
hypersurface defined by a fixed value of t. The point F represents
the two-sphere of radius
$\pi\m{K\,sinh^{(2/\epsilon)}({{\epsilon\,t}\over{2L}}}$) in that
spacelike surface. This two-sphere can be enclosed in a cube of side
length 2$\pi\m{K\,sinh^{(2/\epsilon)}({{\epsilon\,t}\over{2L}}}$),
and this cube can be allowed to expand or contract as we follow the
sphere either into the future or the past along the geodesic OFB. We
now perform the usual identifications of the faces of these cubes,
to obtain tori. The effect on the Penrose diagram is [to a good
approximation] to cut away all parts of the original triangle which
lie to the right of the chosen geodesic OFB; that is, we cut away
the dotted region in Figure 2. Note that conformal infinity is also
compactified: it is represented by the solid part, AB, of the upper
horizontal line, and of course it is also a torus. The singularity
is represented by a single point, O.

Now note that we do \emph{not} expect holography to resolve the
singularity in Figure 1, at least not directly. Instead, the
boundary theory must in some way inform us that values of
$\epsilon$ greater than 2 are not appropriate for describing the
pre-inflationary era. For, as we discussed earlier, the matter
content of those spacetimes is such that gravity is attractive in
the earliest era, and so a singularity is natural. The singularity
should be \emph{directly} resolved only in those cases where the
unusual matter content of the early universe [dominated by
extended objects such as strings] is such as to lead us to expect
that a singularity is not physically reasonable. That is,
holography should work in a way so that the singularity at O in
Figure 2, but \emph{not} the one in Figure 1, is revealed to be
unphysical. How can the boundary physics distinguish the
singularities in Figures 1 and 2?

Consider a timelike geodesic extending to the past from the point D.
In the diagram it appears to ``bounce" off the line OFB, because it
re-enters the torus on the opposite side when it passes beyond the
most distant point from the origin; it then proceeds towards the
line OA [the worldline of the origin], and again appears to
``bounce" as it passes through the origin. It reaches the
singularity at O in a finite proper time, but it must circumnavigate
the torus \emph{a literally infinite number of times} in order to do
this. It is not entirely clear what this statement means, and it is
particularly confusing if we recall that circumnavigations can have
the effect of some discrete transformation such as charge
conjugation, so that this transformation is being applied \emph{an
infinite number of times in a finite period of proper time}. It
follows, for example, that it simply does not make sense to ask
whether a given object, such as a quark, was initially a quark or an
anti-quark. We find ourselves getting entangled in the traditional
paradoxes associated with infinity.

This is the classical way of stating the case. In reality, quantum
effects would render this situation unphysical well before the
singularity is reached: quantum uncertainties, combined with the
effects of parallel transport, would make it impossible to specify
whether a given object is a particle or an anti-particle for some
interval of time after the singularity. The nature of the system and
the physics of its interactions cannot be properly formulated under
such conditions.

These paradoxes simply indicate that we are dealing with a
procedure that cannot actually be realized: that is, the
singularities in the metrics with $\epsilon\;\leq\;2$ cannot
really be attained by any actual particle or antiparticle, and
they should be regarded as a mathematical fiction. The question
then is: what takes the place of the region near O in Figure 2?

The most straightforward assumption is that the paradoxes we have
been discussing are avoided because the spatial sections of the
Universe are never in fact smaller than a certain size; this is
also compatible with the ideas of string gas cosmology
\cite{kn:bat}. That is, the Universe must be born along some
spacelike hypersurface large enough so that the status of a given
object is always well-defined modulo a \emph{finite} number of
discrete transformations\footnote{Alternatively, the Universe
could be asymptotic in the past to some quasi-static state, as in
\cite{kn:soo}.}.

This is just the ``creation from nothing" scenario
\cite{kn:vilenkin1}\cite{kn:vilenkin2}, recently revived and
improved in various ways by Tye and co-workers
\cite{kn:tye}\cite{kn:sarangi}\cite{kn:sashtye} and by Ooguri et
al \cite{kn:ooguri}; see also \cite{kn:laura1} for similar
approaches. However, if we wish to avoid violating the Null Energy
Condition, it is difficult to arrange for the Universe to be
created smoothly \emph{along a three-torus}; in fact, it can only
be done by modifying the Einstein equations \cite{kn:OVV}. A
simple way of performing such a modification, using the ideas of
Gabadadze and Shang \cite{kn:gab1}\cite{kn:gab2}, was explained in
\cite{kn:singularstable}; the resulting ``Creation From Nothing on
a Torus" metric, defined for t $\geq$ 0, is given by
\begin{equation}\label{eq:K5}
g\m{_{CFNT}(K,\,L) \;=\; dt^2\; -\;
K^2\;cosh^{(2/3)}\Big({{3\,t}\over{L}}\Big)\,[d\theta_1^2 \;+\;
d\theta_2^2 \;+\; d\theta_3^2]}.
\end{equation}
It is completely non-singular and smooth.

In this roundabout way, we find that the topology of the boundary,
expressed by the effects discovered by Krauss, Wilczek, and
Preskill, does seem to have profound consequences for bulk physics
in the case depicted in Figure 2 [and only in that case]. In fact,
the effects due to non-trivial gauge holonomy become more pronounced
as we penetrate deeper into the bulk, that is, farther into the past
and closer to the apparent initial singularity. By requiring that
the bulk geometry should account for the boundary gauge holonomy
group in a reasonable way, we may be able to resolve the singularity
portrayed in Figure 2. One might call this an example of
\emph{topological holography}.

Obviously this is a mere sketch of a novel way in which dS/CFT might
guide us towards a resolution of the initial singularity. What this
discussion really tells us is that as we work towards a better
understanding of de Sitter holography, we should be aware of the
possibility that the gauge group may be disconnected or discrete,
since this may well have profound consequences both for the boundary
field theory and for the bulk.

\addtocounter{section}{1}
\section*{\large{\textsf{5. Conclusion }}}

We have seen that, once its limitations are understood, the dS/CFT
correspondence may have much to teach us about cosmology. We found
that the apparently unreasonable Strominger bound is seen to be
perfectly reasonable when it is related to the equation of state of
cosmological matter. We then asked whether dS/CFT can have any
bearing on the problem of the initial singularity in cosmology: can
a gauge theory at future infinity influence the physics of the deep
bulk? While we were not able to answer this question definitively,
we have uncovered some hints suggesting that the answer may be
positive.

The specific examples of gauge theories and spacetimes we have
discussed here are of course over-simplified. They should be
regarded as suggestions as to the lines along which one might work
when the dS/CFT correspondence is better understood. What this
discussion does clearly show is that the \emph{topology} of the
future conformal boundary has a strong influence \emph{both} on the
gauge theory at infinity \emph{and} on bulk physics. On one side,
non-trivial topology leads to unusual physics due to holonomy
effects, while on the other it may well enforce a modification of
the Einstein equations, leading to a replacement of the singularity
shown in Figure 2 by ``creation from nothing". Clarifying the
connections between these two ideas may help us to understand both
the dS/CFT correspondence and the problem of the initial
singularity.

\addtocounter{section}{1}
\section*{\large{\textsf{ Acknowledgements}}}
The author is extremely grateful to Soon Wanmei for preparing the
diagrams and for many helpful discussions.

\end{document}